# Bifurcation analysis of strongly nonlinear injection locked spin torque oscillators

J. Hem[1], L.D. Buda-Prejbeanu[1] and U. Ebels[1]

[1] Univ. Grenoble Alpes, CEA, CNRS, Grenoble INP, IRIG-Spintec, 38000 Grenoble, France

**Abstract.**

We investigate the dynamics of an injection locked in-plane uniform spin torque oscillator for several forcing configurations at large driving amplitudes. For the analysis, the spin wave amplitude equation is used to reduce the dynamics to a general oscillator equation in which the forcing is a complex valued function $\mathcal{F}(p,\psi) \propto \varepsilon_1(p)cos(\psi) + i\varepsilon_2(p)sin(\psi)$. Assuming that the oscillator is strongly nonisochronous and/or forced by a power forcing ($|\nu\varepsilon_1/\varepsilon_2| \gg 1$), we show that the parameters $\varepsilon_{1,2}(p)$ govern the main bifurcation features of the Arnold tongue diagram : (i) the locking range asymmetry is mainly controlled by $d\varepsilon_1(p)/dp$, (ii) the Taken-Bogdanov bifurcation occurs for a power threshold depending on $\varepsilon_{1,2}(p)$ and (iii) the frequency hysteretic range is related to the transient regime through the resonant frequency at zero mismatch frequency. Then, the model is compared with the macrospin simulation for driving amplitudes as large as $10^0 - 10^3 A/m$ for the magnetic field and $10^{10} - 10^{12} A/m^2$ for the current density. As predicted by the model, the forcing configuration (nature of the driving signal, applied direction, the harmonic orders) affects substantially the oscillator dynamic. However, some discrepancies are observed. In particular, the prediction of the frequency and power locking range boundaries may be misestimated if the hysteretic boundaries are of same magnitude order. Moreover, the misestimation can be of two different types according if the bifurcation is Saddle node or Taken Bogdanov. These effects are a further manifestation of the complexity of the dynamics in nonisochronous auto-oscillators.

## 1. INTRODUCTION

Spin-torque nano-oscillators (STNOs) [1] have appealing synchronization regime thanks to their unique ability to lock under many types of driving signals [2,3,4]. For this, they are promising building blocks for future microwave devices, such as frequency dividers [5], energy efficient frequency detectors [6] or unconventional computing architectures [7,8]. Besides, the synchronization of STNOs continues to challenge our understanding of the magnetization dynamics [9,10,11]. One major issue of the injection locking regime, between an oscillator and a generator, is the prediction of the dynamics as the driving amplitude increases. The locking range undergoes complex asymmetrical and hysteresis effects, due to the nonlinear (i.e. nonisochronous) nature of STNOs (free running frequency dependance with the supply current) and whose understanding is of practical importance since they even occur at experimentally achievable driving amplitude [12,13,14,15].

The dynamics of the injection-locked STNO can be addressed by the direct study of the Landau - Lifshitz - Gilbert - Slonczewski (LLGS) equation. Numerical and analytical studies [16,17,18] demonstrate the existence of different types of stable equilibria in the system. The oscillator and the generator can be either synchronized ("P region" as originally labeled by Ref. [16]) or unsynchronized ("Q region") depending on the frequency mismatch between them. Moreover, a metastable region may be created for sufficiently high driving amplitudes and frequency mismatch where both synchronized and unsynchronized states coexist ("P/Q region"), leading to the locking range hysteresis. The determination of the bifurcation lines shaping these regions in the Arnold tongue diagram (driving amplitude vs. frequency mismatch, Fig. 1) is one major issue. There are three important features, common to the dynamics of STNOs: (i) *Asymmetry of the locking range:* the locking range boundaries arising from a saddle-node bifurcation (red line, Fig. 1) expand non-linearly with the driving amplitude, while shifting towards positive or negative frequency mismatch [19]. This contrasts with the linear and symmetrical boundaries predicted by Adler-like models [16,20,21,22]. (ii) *Partial restriction of the locking range:* for sufficiently large driving amplitudes, one locking range boundary can be limited to a constant frequency [23] (red line, Fig. 1). This desynchronization process results from a Taken-Bogdanov bifurcation [24], in which a stable equilibrium is transformed in an unstable one, via a Hopf bifurcation. (iii) *Hysteresis effect:* hysteretic boundaries (blue line Fig. 1) also expand non-linearly with the driving amplitude. Although they have been mainly studied by simulations [18,19,25], they can be estimated by Melnikov technics applied to the LLGS equation [16] in case of a highly symmetrical STNO and forcing.

Until now, these bifurcation processes have been studied in case of specific forcings and range of driving amplitudes. However, many others driving conditions exist in STNOs, potentially useful for applications and it is not clear to what extend these properties continue to hold. For these reasons, we search for a way to expand the bifurcation analysis in case of an arbitrary forcing.

In this article, we use the Hamiltonian spin wave formalism [22,26,27,28] to reduce the oscillator dynamics to a general oscillator equation and show that the forcing function $\mathcal{F} = a_g(\varepsilon_1 \cos(\psi-\theta) + i\varepsilon_2 \sin(\psi-\theta))$ through the parameters $\varepsilon_1$, $\varepsilon_2$ is fully related to the bifurcation processes mentioned above - (i) the locking range asymmetry, (ii) the Taken-Bogdanov bifurcation and (iii) the hysteresis. Then, the calculations are applied and confronted with the LLGS macrospin simulations to investigate the dynamics of different forcing configurations of a synchronized in-plane uniform STNO. The forcings differ from the nature of the driving signals (magnetic field or spin polarized current), the applied direction, or the harmonic frequencies. For the analysis, we have paid attention to study the regime of low as well as large driving amplitudes, for which experiments, or applications may be carried out. Moreover, it is important to note that the STNO studied here is highly nonisochronous $|\nu| \gg 1$, with $\nu$ the normalized non-linear shift factor [27,28]. Although the nonisochronicity is often overlooked in standard auto-oscillator analysis, at most viewed as a perturbation [22,29], the situation may be reversed for STNOs. For the calculations, we have preserved the strong nonisochronicity of the system by the assumption $|\nu\varepsilon_1/\varepsilon_2| \gg 1$ and have shown that the extraction of the Arnold tongue diagram considerably simplifies, while being valid for a large range of driving conditions.

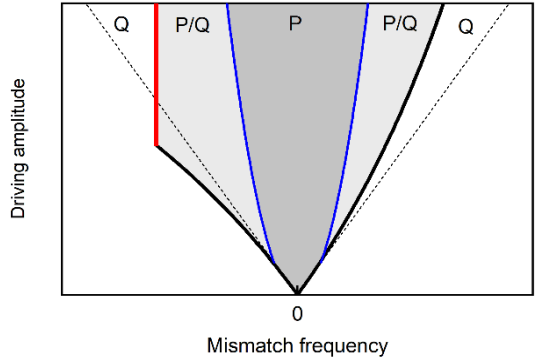

*Fig 1. Typical Arnold tongue diagram of an injection locked spin torque oscillator. "P", "Q" and "P/Q" regions refers to the type of stable states in the system described in Ref. [16]. Locking range boundaries, arising from saddle-node bifurcation (black line) and from Taken-Bogdanov bifurcation (red line). Hysteretic boundaries (blue line). Locking range boundaries predicted by Adler-like models (Dashed line).*

The article is organized as follows. In Part. 2, we describe the uniform spin-torque oscillator and the forcings investigated. The LLGS equation and the oscillator equations are presented. Then in the successive parts, we relate the bifurcation process of the Arnold tongue diagram with the form of the forcing and apply the results to several concrete cases. Part. 3 discusses the asymmetric behavior of the locking range where the role of $d\varepsilon_1/dp$ is emphasized. Part. 4 is dedicated to the Taken-Bogdanov (TB) bifurcation and shows the important role of the oscillation power and the forcing. Part. 5 addresses the hysteresis effect in which a general relation connecting the frequency hysteretic boundaries and the transient regime of the oscillator is proposed. Finally, Part. 6 presents the Arnold Tongue diagrams of the forcings to summarize the results and confront the model with the simulation in the limit of high driving amplitude. Some discrepancies exist when the locking range boundaries increase drastically or if the hysteretic bifurcation bounds are of same order of magnitude as the locking range.

## 2. DESCRIPTION OF THE OSCILLATOR

The STNO considered here is a uniformly in-plane (IP) magnetized magneto-resistive nanopillar device with an elliptical cross-section, called for short IP-STNO, Fig. 2a. The free layer magnetization $\boldsymbol{m} = (m_x, m_y, m_z)$ follows the LLGS equation [30] given by:

$$\dot{\boldsymbol{m}} = -\gamma_o'(\boldsymbol{m} \times \boldsymbol{H}_{eff}) - \gamma_o'\alpha\, \boldsymbol{m} \times (\boldsymbol{m} \times \boldsymbol{H}_{eff}) - \gamma_o' a_{J0} J_{dc}\, \boldsymbol{m} \times (\boldsymbol{m} \times \boldsymbol{m}_p) + (\dot{\boldsymbol{m}})_F(t) \quad (1)$$

With $\gamma_o' = \gamma_0/(1+\alpha^2)$ the modified gyromagnetic ratio. The effective magnetic field $\boldsymbol{H}_{eff} = \boldsymbol{H}_{dc} + \boldsymbol{H}_{dem}$, defines the equilibrium direction $\boldsymbol{u}_x$ of the system, with the static applied field $\boldsymbol{H}_{dc} = H_{dc}\boldsymbol{u}_x$, the demagnetizing field $\boldsymbol{H}_{dem} = -M_s \overline{\overline{\boldsymbol{N}}}\boldsymbol{m}$ and the saturation magnetization $M_s$. The diagonalized demagnetizing tensor $\overline{\overline{\boldsymbol{N}}}$ with $N_{zz} \gg N_{yy} > N_{xx}$, causes both the strong elliptical motion of $\boldsymbol{m}$ around the $x$-axis and the nonisochronous nature of the oscillator. The constant $\alpha$ is the Gilbert damping, $J_{dc}$ is the $dc$ current density, spin polarized in-plane into the direction $\boldsymbol{m}_p = (m_{px}, m_{py}, 0)$ with $a_{J0}$ the pre-factor. For simplicity, the field-like torque is neglected. The last term is the time dependent forcing torque $(\dot{\boldsymbol{m}})_F(t)$ that arises from an $rf$ applied external generator signal $s(t) = a_g \cos(\omega_g t)$ with $a_g > 0$ the amplitude and $\omega_g$ the microwave angular frequency.

For the numerical evaluation the operational conditions are: $\gamma_0 = 2.210 \times 10^5\, m.A^{-1}.s^{-1}$, $H_{dc} = 3.183 \times 10^4\, A.m^{-1}$, $M_s = 1 \times 10^6 A.m^{-1}$, $N_{zz} = 0.8880$, $N_{yy} = 0.0594$, $N_{xx} = 0.0525$ for the conservative part and $\alpha = 0.02$, $\boldsymbol{m}_p = (m_{px}, m_{py}, 0)$ with $(\boldsymbol{m}_p, \boldsymbol{u}_x) = 165°$, the $a_{J0} = -2.501 \times 10^{-8}\, m$ for the dissipative part. Only the amplitude $J_{dc}$ of the dc current density and the forcing torque are considered as variable.

The general oscillator equation for this IP-STNO, in terms of the complex variable $d = \sqrt{p}\, exp(i\psi/n)$, is derived from the LLGS equation Eq. 1 using the Hamiltonian formalism for spin waves [26,27,22]:

$$\dot{d} = i(\omega(p) - \omega_g/n)d - \Gamma(p)d + \mathcal{F}(p,\psi)d \quad (2a)$$

The equation rewrites for real variables $(p, \psi)$ as:

$$\dot{p} = 2p\left(-\Gamma(p) + a_g \varepsilon_1(p)\cos(\psi - \theta)\right) \quad (2b)$$

$$\dot{\psi} = n\omega(p) - \omega_g + a_g n\varepsilon_2(p)\sin(\psi - \theta) \quad (2c)$$

Eq. 2(a-c) express the motion of the free layer magnetization in the rotating frame of the generator, with $p = |d|^2$ the oscillation power, $\psi = n\phi - \omega_g t$ the phase difference between $n$ times $\phi$ the STNO phase and $\omega_g t$ the generator phase, $\omega(p)$ the STNO angular frequency and $n$ the order of synchronization. Although the shift angle $\theta$ is important as it contributes to a supplementary phase difference shift in the locked state, it has no influence on the bifurcation properties so that it will be neglected in the following ($\theta = 0$). The precession torque induces the power dependence of the oscillator frequency given by the relation $\omega(p) = \omega(0) + \omega_{NL}(p)$. The dissipative torques are related to the damping function $\Gamma(p) = \Gamma_+(p) - \Gamma_-(p)$ with the natural damping $\Gamma_+(p)$ and the current induced damping $\Gamma_-(p)$.

Applied to the IP-STNO considered here under the given operational conditions, the spin wave transformation leads to the following values and relations: $\omega(0)/(2\pi) = 6.441\ GHz$, $\omega_{NL}(p) = Np$ with $N/(2\pi) = -5.030\ GHz$, $\Gamma_+(p) = \Gamma_G(1 + Q_1 p)$ with $\Gamma_G/(2\pi) = 318\ MHz$, $Q_1 = -0.350$ and $\Gamma_-(p) = \sigma J_{dc}(1-p)$ with $\sigma = -5.337 \times 10^{-3}\ A^{-1} \cdot s^{-1} \cdot m^2$. The critical current density above which stable limit cycles are stabilized is $J_{cr} = \Gamma_G/\sigma = -375 \times 10^9\ A \cdot m^{-2}$ and the free running power $p_0$ is related to $J_{dc}$ through $p_0 = (J_{dc}/J_{cr} - 1)/(J_{dc}/J_{cr} + Q_1)$, Fig. 2b. Further details can be found in Ref. [28]. As an example, for $J_{dc} = -400 \times 10^9\ A \cdot m^{-2}$ or $p_0 = 0.093$, the value for the free running frequency is $\omega(p_0)/2\pi = 5.974\ GHz$, the damping rate for small power deviation is $\Gamma_p(p_0)/(2\pi) = 21\ MHz$ with $\Gamma_p(p_0) = p_0 d\Gamma(p_0)/dp$ and the normalized non-linear shift factor is $\nu(p_0) = Np_0/\Gamma_p(p_0) = -22$. Note that for the IP-STNO oscillator $N < 0$, thus the free running frequency decreases as the oscillation power increases. Furthermore $|\nu| \gg 1$ which is typical of a strongly nonisochronous oscillator.

The forcing function in Eq.2 is, in its most general form, given by $\mathcal{F}(p,\psi) = a_g(\varepsilon_1(p)\cos(\psi) + i\varepsilon_2(p)\sin(\psi))$, where $\varepsilon_1, \varepsilon_2$ are the real-valued coupling terms that describe the coupling between the STNO oscillation mode and the rf source. They depend on the nature, the direction $\boldsymbol{u}$, and the harmonic order of the forcing torque $(\dot{\boldsymbol{m}})_F$, leading to different power dependencies of the coupling terms $\varepsilon_1(p) > 0, \varepsilon_2(p)$. These have been calculated in Ref. [28] for different basic forcing configurations. To demonstrate the non-linear effects of injection locking, three different configurations are considered here, that are referred to as "$Dx2$", "$Cx2$" and "$Cy1$" (letters $C$ and $D$ recall the conservative or dissipative nature of the forcing torque, $x$ or $y$, the direction of applied field or spin polarization and $n$ the number of the harmonic order).

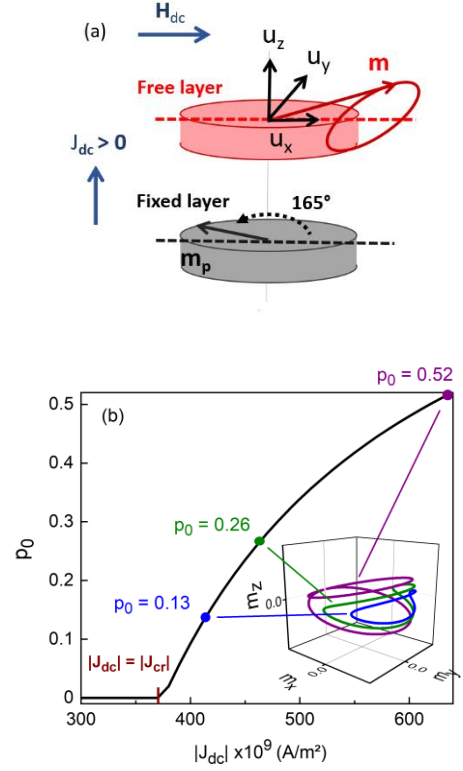

*Fig. 2. (a) Schematic of the in-plane uniform spin-torque oscillator. Magnetization trajectories are elliptical around the x-axis with a small out-of-plane component $m_z$. The elliptical shape is captured by the spin wave transformation through the elliptical parameter $-1 < \mathcal{B}/\mathcal{A} < 1$ with here $\mathcal{B}/\mathcal{A} = -0.91$. Full relations between the power $p$ and the components $m_x, m_y, m_z$ are given in Appendix A. (b) Analytical relation between the free running power $p_0$ and the applied dc current density $J_{dc}$ for the full range of the in plane precession mode $(0 < p_0 < 0.52)$.*

Case "Dx2", with $(\dot{\boldsymbol{m}})_F = -\gamma'_o J_g a_{J0} m_{px} \cos(\omega_g t) \boldsymbol{m} \times (\boldsymbol{m} \times \boldsymbol{u_x})$ represents a dissipative driving force due to an rf current density $J_g \cos(\omega_g t)$ with the spin polarization component $m_{px}$ along $\boldsymbol{u_x}$, at the 2$^{nd}$ harmonic order $n = 2$. The corresponding forcing function $\mathcal{F}$ is a pure power forcing function with $\varepsilon_1(p) = |\mathcal{B}/\mathcal{A}|p$, $\varepsilon_2 = 0$ and $a_g = -\gamma'_o a_{J0} m_{px} J_g/2$. Note that $\varepsilon_1(p)$ depends linearly on the power $p$, as well as on the ellipticity $\mathcal{B}/\mathcal{A}$ of the STNO trajectory, see Ref. [28].

Case "Cx2", with $(\dot{\boldsymbol{m}})_F = -\gamma'_o H_g \cos(\omega_g t) \boldsymbol{m} \times \boldsymbol{u_x}$ describes a conservative driving force due to an rf magnetic field $\boldsymbol{H_g} = H_g \cos(\omega_g t) \boldsymbol{u_x}$ at $n = 2$ applied parallel to the equilibrium direction $\boldsymbol{u_x}$. In contrast to the case of "Dx2", the corresponding forcing function $\mathcal{F}$ has a circular form with equal strengths of the coupling terms $\varepsilon_1 = -\varepsilon_2 = |\mathcal{B}/\mathcal{A}|/\sqrt{1 - (\mathcal{B}/\mathcal{A})^2}$ and with $a_g = -\gamma'_o H_g/2$. In this case, $\varepsilon_1, \varepsilon_2$ are independent of the oscillation power $p$, but depend on the ellipticity $\mathcal{B}/\mathcal{A}$.

Case "Cy1", with $(\dot{\boldsymbol{m}})_F = -\gamma'_o H_g \cos(\omega_g t) \boldsymbol{m} \times \boldsymbol{u_y}$ describes a driving force due to an rf magnetic field $\boldsymbol{H_g} =$ $H_g cos(\omega_g t)\boldsymbol{u}_y$ at $n = 1$ applied parallel to $\boldsymbol{u}_y$ (in-plane and perpendicular to the equilibrium direction). The coupling terms in $\mathcal{F}$ are unequal $\varepsilon_1(p) \neq \varepsilon_2(p)$ with a higher order dependence on the oscillation power (see Ref. [28] for the full expressions of $\varepsilon_1$ and $\varepsilon_2$) and with $a_g = -\gamma'_o H_g/2$.

In the next parts, we shall see that these specific forcings can illustrate many of the different bifurcation processes allowed by the injection locked STNO, thanks to all their differences in the coupling terms.

## 3. ASYMETRICAL LOCKING RANGE

3.A General stationary solutions

As a preliminary, we describe the general approach to determine the asymmetric locking boundaries from the stationary solutions of Eq. 2(a-c) ($\dot{\psi} = 0, \dot{p} = 0$). To simplify the calculations, we assume $|\nu\varepsilon_1/\varepsilon_2| \gg 1$, which is verified for a large range of STNOs (such as the IP-STNO considered here) because they are strongly nonlinear $|\nu| \gg 1$ and the forcings verify $|\varepsilon_1/\varepsilon_2| \geq 1$. Consequently, the phase forcing component $\varepsilon_2 sin(\psi)$ can be neglected in Eq. 2c. From this one obtains directly the relation between the STNO power $p$ and STNO frequency $\omega(p) = \omega(0) + \omega_{NL}(p) = \omega_g/n$ in the injection locked state which has the same dependence as the power $p_0$ and frequency $\omega(p_0) = \omega(0) + \omega_{NL}(p_0)$ in the free running state. With this, the stationary power of the locked state can be determined from the power equation Eq. 2b as: $\varepsilon_1(p) \cos(\psi) = \Gamma(p)$. Moreover, the locking range boundaries are defined by $cos(\psi) = \pm 1$. Then if the power $p = p_0 + \Delta p$ is shifted by an amount $\Delta p$ from the free running power $p_0$ in the injection locked state, one can expand Eq. 2(b,c) around $p_0$:

$$a_g \varepsilon_1(p_0) + \sum_{k \geq 1} \left(a_g \varepsilon_1^{(k)}(p_0) \pm \Gamma^{(k)}(p_0)\right) \frac{\Delta p_\pm^k}{k!} = 0 \quad (3a)$$

$$\delta_\pm = \sum_{k \geq 1} n \omega_{NL}^{(k)}(p_0) \frac{\Delta p_\pm^k}{k!} \quad (3b)$$

In Eq. 3(a,b) the terms $x^{(k)}(p_0)$ are the k-th order derivative with respect to the power of the functions $x(p)$, that are evaluated at the free running power $p_0$ and the free running frequency mismatch defined by $\delta(p) = \omega_g - n\omega(p)$. Eq. 3a provides analytical expressions for the maximum power shifts $\Delta p_\pm$ ($\Delta p_- < 0 < \Delta p_+$) at the upper and lower locking boundaries, and by inserting into Eq. 3b also the maximum frequency mismatches $\delta_\pm$ ($\delta_- < 0 < \delta_+$).

In the lowest order, the symmetric locking boundaries $\delta_+ = -\delta_-$ are determined from Eq. 3 by $\varepsilon_1(p_0)$ and by the first order derivatives $\Gamma^{(1)}(p_0) = \Gamma_p(p_0)/p_0$, $\omega_{NL}^{(1)}(p_0) = N$, with $\Gamma_p(p_0)$ the amplitude relaxation rate [16,22]. It leads to the corresponding symmetric locking range:

$$\Delta\Omega_S = \frac{\delta_+ - \delta_-}{2} = n|\nu(p_0)|\varepsilon_1(p_0)a_g \quad (4)$$

with the proportionality factor $|\nu(p_0)| \approx \sqrt{1 + \nu(p_0)^2}$ due to the non-linear amplitude-phase coupling [16,22]. However for larger driving amplitudes $a_g$, the higher order terms $\Gamma^{(k)}(p_0)\Delta p$, $\omega_{NL}^{(k)}(p_0)\Delta p$ ($k \geq 2$) and $\varepsilon_1^{(k)}(p_0)\Delta p$ ($k \geq 1$) cannot be neglected anymore, leading to asymmetrical boundaries with $|\delta_+| \neq |\delta_-|$.

In particular, we show next that asymmetry arises if at least the first order term of the forcing $\varepsilon_1^{(1)}(p_0)\Delta p$ is included. In case of non-zero values of the first orders $\Gamma^{(1)}(p_0) = \Gamma_p(p_0)/p_0$, $\omega_{NL}^{(1)}(p_0) = N$ and $\varepsilon_1^{(1)}(p_0)$ and zero values for higher orders, the stationary solutions of Eq. 2(a-c) yield the power shift $\Delta p$ as a function of the frequency mismatch $\delta = \omega_g - n\omega(p_0)$ as well as the phase difference $\psi$ [31]:

$$\Delta p = \delta/(nN) \quad (5a)$$

$$\psi = arcsin\left(\frac{\Gamma_p \Delta p}{a_g p_0 (\varepsilon_1 + \varepsilon_1^{(1)} \Delta p)}\right) + \pi/2 \quad (5b)$$

From Eq. 3 or Eq. 5 the maximum power shifts $\Delta p_\pm$ are obtained and subsequently the maximum frequency mismatches $\delta_\pm$, see Eq. 6:

$$\Delta p_\pm = \frac{\pm a_g p_0 \varepsilon_1(p_0)}{\Gamma_p(1 \mp A_s)} \quad (6a)$$

$$\delta_\pm = \frac{\mp Sign(N)\Delta\Omega_S}{1 \pm A_s} \quad (6b)$$

$$A_s = a_g \varepsilon_1^{(1)}(p_0) \frac{p_0}{\Gamma_p(p_0)} \quad (6c)$$

With $\Delta\Omega_S$ the symmetric locking range defined in Eq. 4. The parameter $A_s$ can be viewed as an asymmetrical rate since:

$$\frac{\delta_- + \delta_+}{\delta_- - \delta_+} = A_s \quad (7)$$

In Eq. 6, we see that the maximum power shifts $\Delta p_\pm$ and maximum frequency mismatches $\delta_\pm$ depend on the forcing through the driving amplitude $a_g$ and on the zero and first order coupling terms $\varepsilon_1(p_0)$ and $\varepsilon_1^{(1)}(p_0)$. The asymmetry of the locking range is captured by the parameter $A_s$, which also depends on $a_g$ and $\varepsilon_1^{(1)}(p_0)$. One can distinguish two different cases:

$A_s \neq 0$, *asymmetrical case:* This situation occurs for non-zero values of the driving amplitude $a_g$ and non-zero $\varepsilon_1^{(1)}(p)$. In particular, the asymmetry parameter $|A_s|$ increases linearly with $a_g$ and $|\varepsilon_1^{(1)}(p)|$. Regarding the asymmetry of the power shift $\Delta p_\pm$, it depends on the sign of $A_s$ and exclusively on $\varepsilon_1^{(1)}$. For example, if $\varepsilon_1^{(1)} > 0$ (resp. $\varepsilon_1^{(1)} < 0$), the forcing strength increases (decreases) with the oscillation power, such that the upper bound of the power shift $|\Delta p_+|$ will be larger (smaller) than the lower bound of the power shift $|\Delta p_-|$. For the asymmetry of $\delta_\pm$, one needs also to consider the sign of the non linearity $N$ or $\nu$. For example, for $N < 0$ (as is the case of the IP-STNO), one should expect that if $\varepsilon_1^{(1)} > 0$, the negative frequency bound $|\delta_-|$ will be larger than the positive one $|\delta_+|$ and the inverse if $N > 0$.

$A_s = 0$, *symmetrical case:* Exact or quasi symmetrical locking range is possible in two cases: (i) If the driving amplitude $a_g$ is small for whatever value of $\varepsilon_1^{(1)}$. In other words, Adler-like models can be applied in any situation in the limit of small driving amplitude which is in agreement with Ref. [22]. From the asymmetry parameter $A_s < 1$ one can define an upper limit for the linear range as $a_g < \frac{\Gamma_p(p_0)}{\varepsilon_1^{(1)}(p_0)p_0}$, that is given through $p_0$ by the operational point, the corresponding amplitude relaxation rate $\Gamma_p$ and the coupling terms $\varepsilon_1^{(1)}$. (ii) For any values of the driving amplitude, provided that the forcing does not vary much with the oscillation power, i.e. for $\varepsilon_1^{(1)}(p) \approx 0$. Thus, the derived equations Eq. 6 predict that a symmetric locking range is possible also at higher driving amplitude.

As a concluding note, it is mentioned that the results of Part. 3A can be readily applied to discuss the asymmetry in a wide range of uniform STNOs and forcings with the full expressions of $\varepsilon_1, \varepsilon_2$ given in Ref. [28], which is useful to explore any other configurations.

3.B Applications to the IP-STNO

The development presented above is now applied to the IP-STNO. Since at high driving amplitude the coupling term $\varepsilon_1(p)$ and its derivative $\varepsilon_1^{(1)}(p_0)$ mainly govern the locking range behavior and the asymmetry of its boundaries $\Delta p_\pm$ and $\delta_\pm$, see Eq. 6, the power dependencies for the three forcing functions, "*Dx2*", "*Cx2*", "*Cy1*" are shown in Fig. 3. Their quite different dependencies predict different types of asymmetries. In Fig. 3a, we see that for the forcing "*Dx2*", $\varepsilon_1(p)$ increases with $p$ ($\varepsilon_1^{(1)}(p) > 0$) whereas in the case of the forcing "*Cy1*", $\varepsilon_1(p)$ decreases with $p$ ($\varepsilon_1^{(1)}(p) < 0$). As a result, the asymmetry $A_s$ has a different sign, so that the locking range, centered at a given free running power $p_0$, will have an opposite direction of asymmetry. Unlike the two, the forcing case "*Cx2*" verifies $\varepsilon_1^{(1)}(p) = 0$ and therefore no asymmetry is expected in this case. Moreover in Fig. 3b, we see that $|\varepsilon_1^{(1)}(p_0)|$ may or not depend on $p_0$ so that it is also the case for the strength of the asymmetry. For example, for the forcing "*Cy1*", the asymmetry decreases with $p_0$ since $|\varepsilon_1^{(1)}(p_0)|$ also decreases with $p_0$, whereas for the forcing "*Dx2*", the asymmetry is independent on $p_0$.

Besides the power dependence, the asymmetry $A_s$ may also depend on the ellipticity $\mathcal{B}/\mathcal{A}$ of the STNO trajectory, through $\varepsilon_1(p)$. For instance, the forcing case "*Cx2*" is the only one whose $\varepsilon_1^{(1)}(p)$ remains zero whatever the ellipticity $\mathcal{B}/\mathcal{A}$ so that the locking range will remain symmetrical whatever the oscillator configuration. This contrasts with the forcing case "*Dx2*" for which $|\varepsilon_1^{(1)}(p)| = |\mathcal{B}/\mathcal{A}|$ where an increase of $\mathcal{B}/\mathcal{A}$ will increase the asymmetry. A similar dependence is expected for the forcing case "*Cy1*". More generally, we expect from the calculation of Ref. [28], that a strong ellipticity ($|\mathcal{B}/\mathcal{A}|\sim 1$) such as for the IP-STNO, strengthens the asymmetry, in contrast to other STNO configurations with more circular trajectories ($\mathcal{B}/\mathcal{A} = 0$), supporting out-of-plane oscillation modes.

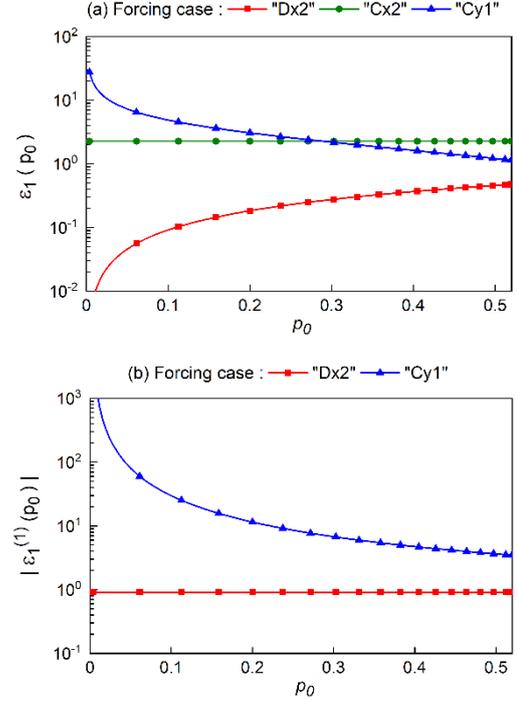

*Fig. 3. Theoretical evolution of (a) the power forcing component $\varepsilon_1(p)$ and (b) its first derivative $\varepsilon_1^{(1)}(p)$ with the free running oscillation power $p_0$ and the forcing cases "Dx2", "Cx2", "Cy1" defined in Part 2. Formulas given in Ref.* [28] *with $\mathcal{B}/\mathcal{A} = -0.91$. Note the log scale in the y-axis.*

3.C Macrospin simulations

To further verify the theoretical predictions, we have integrated Eq. 1 with a 2nd order predicator step scheme of time integration 10 fs and compared the results. We start by giving some details about the numerical procedure for synchronization and how the data are extracted, see Ref. [31] for a full description. To synchronize the STNO, the applied generator frequency $\omega_g$ is swept around $\omega$ the STNO frequency or one of its harmonics by a frequency increment which depends on the desired precision on the locking range value (between 1 to 100MHz). For each increment, the magnetization $\boldsymbol{m}$ of the STNO is stabilized during 400ns and then numerically converted to the variables $p$ and $\psi$ of the model. Note that due to the presence of locking range hysteresis, it is necessary to perform two scans of the frequency mismatch, one in the increasing direction and another in the decreasing direction, in order to access the whole injection locked states defined by $\omega = \omega_g/n$. The procedure is illustrated in Fig.4a which shows the evolution of $\omega$ the STNO frequency depending on the two scan directions of $\delta$ the frequency mismatch for the forcing case "*Dx2*" detailed below. To maintain the synchronization in the metastable region P/Q, we ensure that between each increment, the magnetization **m** and the generator signal are continuous, the frequency increment change being progressive. Then after combining the data obtained in the two directions, it is possible to get the numerical locking range boundaries $\delta_\pm$, the hysteric range frequency boundaries $\eta_\pm$ defined in Fig.4a as well as the stationary solutions $\Delta p, \psi$ in the full locking range, which can be compared to the model.

first striking observation is the pronounced asymmetry of the locking range, characterized by $|\delta_-| > |\delta_+|$ and $|\Delta p_+| > |\Delta p_-|$, Fig. 4b, which agrees with the model Eqs. (5, 6). Indeed, for the chosen ratio $J_g/J_{dc}$, the asymmetry value is $A_s = +20\%$ so that the upper and lower bounds of the power shift $\Delta p_\pm$ verify $|\Delta p_+| > |\Delta p_-|$ and the bounds of the frequency mimatch $|\delta_-| > |\delta_+|$ (since $N<0$). In Fig. 4b, we point out that the power shift $\Delta p$ is linear with the detuning as predicted by Eq. 5a. This confirms the validity of the assumption of $|\nu\varepsilon_1/\varepsilon_2| \gg 1$. Such power dependence can be a useful criterion to identify (numerically or experimentally) configurations that fulfill the conditions of either strongly non-linearity $\nu$ or/and a dominant power forcing. In Fig. 4b, we see that the asymmetry expands (contracts) the phase towards the negative (positive) frequency mismatch, while the phase range remains within $\pi$ rad as in Adler-like models [16,22].

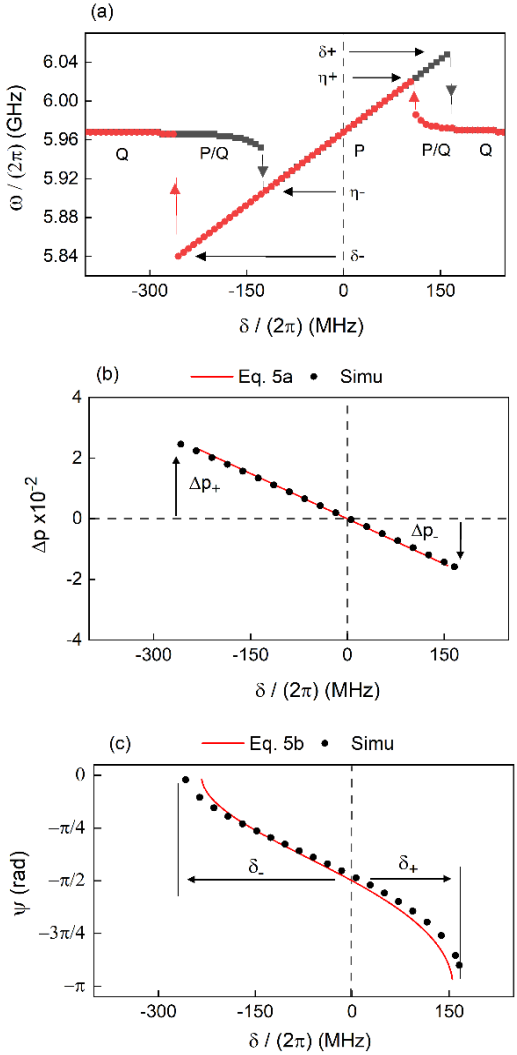

Fig. 4. Forcing case "*Dx2*" at $J_{dc} = -400 \times 10^9 A/m^2$, ($p_0 = 0.093$) applied to the STNO defined in Part. 2. (a) Evolution of $\omega$ the STNO frequency with $\delta$ the freqeucny mismatch between the generator and the STNO, for the increasing (black) and decreasing (red) scan direction of $\delta$. (b) Stationary power shift solution, (c) stationary phase difference solution with the frequency mismatch $\delta$ for $J_g/J_{dc} = 0.3$ and $A_s = +20\%$.

For example, Fig.4(b,c) compare the extracted stationary solutions with the model for the specific case of forcing "*Dx2*" with $J_{dc} = -400 \times 10^9 A/m^2$ and $J_g/J_{dc} = 0.3$. The

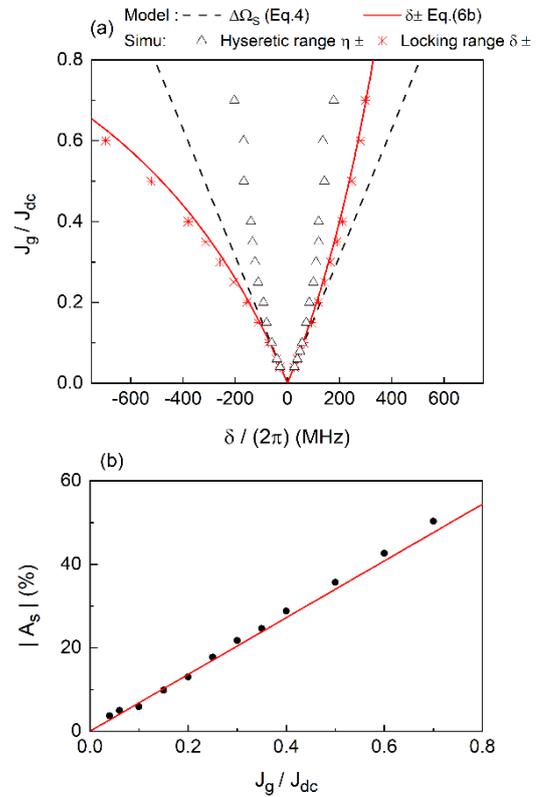

Fig. 5. Forcing case "*Dx2*" at $J_{dc} = -400 \times 10^9 A/m^2$, ($p_0 = 0.093$) applied to the STNO defined in Part. 2. (a) Arnold tongue diagram. Numerical simulation: locking range frequency boundaries $\delta_\pm$ (Red stars), hysteretic range frequency boundaries $\eta_\pm$ (Open triangles). Model: locking range boundaries $\delta_\pm$ Eq.6b (Red line), Symmetric (or Adlerian) locking range boundaries $\Delta\Omega_S$ Eq.4 (Dashed line). (b) Magnitude of the asymmetrical rate $|A_s|$ as a function of the driving amplitude $J_g/J_{dc}$. Numerical simulation obtained from data of (a) with Eq. 7 (black points), Eq. 6c (red line).

The evolution of the locking range boundaries $\delta_\pm$ with the current driving amplitude $J_g/J_{dc}$ is represented in the Arnold tongue diagram Fig. 5a for the specific range of amplitudes: $0 < J_g/J_{dc} < 0.8$. The corresponding evolution of the asymmetry $A_s$ is shown in Fig. 5b. As can be seen, the analytical expressions correctly predict the deviations and curvatures of the numerical boundaries $\delta_\pm$ with the driving amplitude $J_g/J_{dc}$. At small amplitude $J_g/J_{dc} \sim 0.1$, the boundaries $\delta_\pm$ Eq. 6b are fully equivalent to the linear and symmetric boundaries $\Delta\Omega_S$ Eq. 4, see Fig. 5a. This corresponds to an asymmetry $A_s$ close to zero in Fig. 5b. In contrast at higher excitation, the asymmetry $A_s$ progressively increases with the driving amplitude $J_g/J_{dc}$ and this increase is linear with $J_g/J_{dc}$, as predicted by Eq. 6c. In particular, note that for the IP-STNO and this case of forcing, the asymmetric regime occurs for driving amplitudes as low as 0.1 so that this regime should be mainly observed in experiments.

In the next part, we will see that the forcing parameters $\varepsilon_1, \varepsilon_2$ govern another important non-linear property of STNO injection locking at high driving amplitudes

## 4. TAKEN-BOGDANOV BIFURCATION

Eq. 5 defines the stationary states, bounded by two saddle-node bifurcations. However, for high driving amplitudes, part of these states may become unstable when the frequency mismatch increases, inducing a partial restriction of the locking range. This effect has been identified as a Taken-Bogdanov (TB) bifurcation [23,24] and can be viewed as a consequence of the two dimensional nature of the system. In this part, we want to show: (i) that it can also emerge in the amplitude equation Eq. 2, (ii) the key role of the oscillation power and (iii) the relation with the forcing parameters $\varepsilon_1, \varepsilon_2$.

To identify the TB bifurcation, one must perform a stability analysis of the stationary states Eq. 5, see also Appendix B for full details. For this, we write the Jacobian matrix $\mathcal{J}(p,\psi)$ of Eq. 2 at any power $p$ and phase difference $\psi$:

$$\mathcal{J}(p,\psi) =$$
$$\begin{pmatrix} -2\Gamma_p \frac{(2p-p_0)}{p_0} + 2a_g\left(\varepsilon_1 + p\varepsilon_1^{(1)}\right)\cos(\psi) & -2pa_g\varepsilon_1\sin(\psi) \\ nN + na_g\varepsilon_2^{(1)}\sin(\psi) & na_g\varepsilon_2\cos(\psi) \end{pmatrix} \quad (8)$$

For a stationary state, the dependence in $\psi$ can be substituted by $p$ (Eq. 2(b-c) with $\dot{\psi} = 0$, $\dot{p} = 0$). Moreover, the instability condition is given by $Tr(\mathcal{J}) = 0$: the equilibrium is stable if $Tr(\mathcal{J}) < 0$ and unstable if $Tr(\mathcal{J}) > 0$. After some manipulations, the condition can be rewritten in:

$$p_{TB}(p_{TB} - p_0)\frac{\varepsilon_1^{(1)}}{\varepsilon_1} + (p_{TB} - p_0)\frac{n\varepsilon_2}{2\varepsilon_1} - p_{TB} = 0 \quad (9)$$

Eq. 9 shows that the TB bifurcation occurs when the oscillation power $p$ reaches the threshold $p_{TB}$: the oscillation is stable if $p > p_{TB}$ and unstable if $p < p_{TB}$. For example, if $\Delta p_{TB} = p_{TB} - p_0 < \Delta p_- < 0$, the solutions Eq. 5 remains valid, however if not, the range of frequency mismatch becomes limited by the negative power bound $\Delta p_- = \Delta p_{TB}$. Due to the generality of Eq. 2, we emphasize that Eq. 9 is valid for any type of auto-oscillator or forcing function.

Interestingly, the condition Eq. 9 depends only on the oscillation power $p$ and not on the frequency mismatch $\delta$. Nevertheless, it is possible to rephrase it in terms of a frequency mismatch threshold $\delta_{TB}(\Delta p_{TB})$ providing that the relation between $\Delta p$ and $\delta$ is known. On this last point, we can make use of the assumption $|\nu\varepsilon_1/\varepsilon_2| \gg 1$ and find that $\delta_{TB}$ is simply given by $\delta_{TB} = \Delta p_{TB} nN$. For example, in the case $\delta_{TB} > \delta_+ > 0$, the oscillator would be synchronized for all ranges of frequency mismatch allowed by Eq. 5. On the contrary in the case $\delta_+ > \delta_{TB} > 0$, the range of frequency mismatch would be restricted, and the positive frequency bound $\delta_+$ set to $\delta_+ = \delta_{TB}$.

Moreover, we see that the threshold $p_{TB}$ only depends on few parameters, such as the free running power $p_0$, the order $n$ and the parameters of the forcing function via $\varepsilon_1(p_0), \varepsilon_1^{(1)}(p_0), \varepsilon_2(p_0)$, the driving amplitude $a_g$ being excluded. Although $\varepsilon_2$ has been neglected for the determination of the solutions Eq. 5, it totally affects here the value of $p_{TB}$, showing that the full forcing function contributes to the synchronization behavior even under the assumption $|\nu\varepsilon_1/\varepsilon_2| \gg 1$. To find the value of $p_{TB}$, one can solve numerically or analytically Eq. 9, the latter being possible only for simple expressions of the forcing parameters.

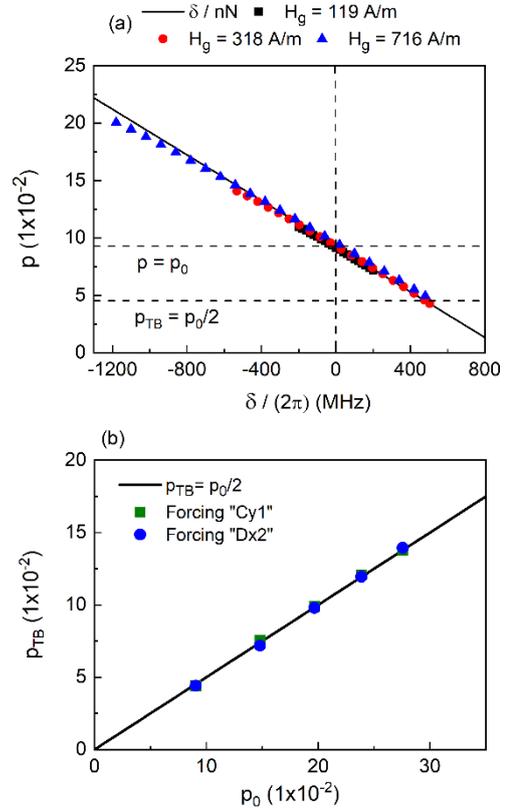

*Fig.6. (a) Stationary power shift solution in case of the forcing "Cx2" at $J_{dc} = -400 \times 10^9 A/m^2$, $(p_0 = 0.093)$ for different amplitude of driving magnetic field $H_g$. (b) Threshold power $p_{TB}$ in function of the free running power $p_0$ in case of the two magnetic field forcings "Cx2" and Cy1".*

For example, the results can be applied to the three forcing of Part. 2. Different power thresholds $p_{TB}$ can be obtained:

(i) *Case "Dx2"*: In this case, the forcing function is a pure power forcing function ($\varepsilon_2 = 0$). Then, Eq. 9 gives two solutions $p_{TB} = 0$ and $p_{TB} = p_0 + \varepsilon_1/\varepsilon_1^{(1)}$. However, since the power forcing component $\varepsilon_1$ verifies $\varepsilon_1 \propto p$, the only remaining solution is $p_{TB} = 0$. In this case, the TB bifurcation occurs at the lowest power possible.

(ii) *Case "Cx2"*: In contrast, the forcing function is a pure circular form and independent of the power ($\varepsilon_1 = -\varepsilon_2$ $\varepsilon_1^{(1)} = 0$). Then, Eq. 9 leads to $p_{TB} = np_0/(n+2)$, where $p_{TB}$ depends only on the free running powers $p_0$ and the order $n$. Note that $p_{TB}$ always verifies $p_{TB} < p_0$ so that the TB bifurcation can always restrict the negative power bound $\Delta p_-$ and respectively the positive frequency bound $\delta_+$ if $N < 0$. With $n = 2$, the power threshold becomes $p_{TB} = p_0/2$.

(ii) *Case "Cy1"*: In this last configuration, Eq. 9 can not be solved analytically as for the two other cases. However, a numerical investigation of Eq. 9 leads to $p_{TB} \approx p_0/2$ and therefore closely the same results as for the forcing *"Cx2"*.

To confront the model with the simulation, we choose to investigate the forcings *"Cx2"* and *"Cy1"* for which $p_{TB} = p_0/2$. Fig 6a shows the evolution of the stationary power solution in case of the forcing *"Cx2"* at several amplitudes of the driving magnetic field $H_g$. We see that the increase of $H_g$ increases the positive power bound $\Delta p_+$ without any restriction as expected. Due to the locking range symmetry ($A_s=0$ case *"Cx2"*), $\Delta p_-$ should decrease as much as the increase of $\Delta p_+$. However, we see for $H_g \geq 320 A/m$, that the decrease of $\Delta p_-$ becomes limited to $\Delta p_- = -p_0/2$. Any larger frequency mismatch causes the desynchronization of the system due to the TB bifurcation. In Fig. 6b, we have checked the dependence of the threshold power $p_{TB}$ on the free running power $p_0$ in case of the two magnetic field forcings *"Cx2"*, *"Cy1"*. The numerical value $p_{TB}$ is reported for each synchronization of the oscillator at different $p_0$. Indeed, we find that $p_{TB}$ varies as $p_{TB} = p_0/2$ as predicted by the model. Therefore, the expansion of the locking range may be intrinsically limited, especially in the case of a forcing function which has non-zero values for the power $\varepsilon_1$ and phase $\varepsilon_2$ forcing components. It is not possible to avoid this effect. However, it is always possible to choose to operate at higher value of $p_0$ in order to increase the corresponding threshold of the frequency mismatch $\delta_{TB}(p_{TB})$ and then limit the potential restriction effect on the locking range.

## 5. HYSTERESIS EFFECT

We address the last important feature which is the emergence of a hysteresis of the locking range in the Arnold tongue diagram. Contrary to the asymmetry or the TB bifurcation, the prediction of the hysteretic boundaries in the diagram is more tedious because it reflects some global properties of the phase diagram, such as the appearance of limit cycles at the same time as stable points. However, we shall see in this part that the hysteretic boundaries are also governed by the forcing parameters. For this, we take advantage of Ref. [16], in which the hysteretic bounds have been estimated by applying the Melnikov technic to the LLGS equation, and adapt the formula to the case of the general oscillator equation for a highly non-linear oscillator. Interestingly, we show that the hysteretic boundaries are related to the transient regime of the oscillator, through the resonant frequency of the stable point at the center of the locking range.

We start by summarizing the hysteretic boundaries given by Eq. 36 in Ref. [16] and found for the case of a highly symmetrical out of plane uniform STNO (i.e. $\mathcal{B}/\mathcal{A} = 0$), synchronized by a circular polarized field at $n = 1$. The formula is valid whatever the oscillator non-linearity. Therefore, if the non-linearity is sufficiently strong, it can be simplified to $\eta_{\pm} \approx \pm \sqrt{[|k_{eff}|h_{a\perp}]}$ with $|k_{eff}|$ the effective anisotropy along the equilibrium axis, $h_{a\perp}$ the amplitude of the driving field and $|k_{eff}| \gg h_{a\perp}$. In terms of the amplitude equation Eq. 2, $|k_{eff}|$ determines the non-linearity $N$ and $h_{a\perp}$ the driving amplitude $a_g$. Regarding the dependence on the oscillation power, although it is neglected in Eq. 36 of Ref. [16], it might be reasonable to assume the form $f(p_0)\varepsilon_1(p_0)$ with an unknown function $f(p_0)$. As a result, a possible expression for the hysteretic boundaries may be:

$$\eta_{\pm} = \pm\sqrt{2n|N|p_0\varepsilon_1(p_0)a_g} = \pm\sqrt{2\Gamma_p\Delta\Omega_S} \qquad (10)$$

With $f(p_0) = 2np_0$ and $\Delta\Omega_S$ the symmetric locking range boundary defined in Eq. 4.

Such phenomenological expression could be applied to a wide range of forcing configurations. Eq. 10 depends on: (i) the driving amplitude $a_g$ which can be of any type, a magnetic field or a spin polarized current, (ii) the harmonic order $n$ and (iii) the free running power $p_0$ via the function $p_0\varepsilon_1(p_0)$. The generality is also illustrated through the second relation, in which the product $\Gamma_p\Delta\Omega_S$ states that hysteresis raises up as soon as there is synchronization.

To verify Eq. 10, we have compared the numerical hysteretic frequency boundaries $\eta_{\pm}$ given by the macrospin simulation and defined in Fig. 4a, with the predicted results, in case of the three forcings *"Cx2"*, *"Cy1"* and *"Dx2"*. The results are plotted in the rescaled Arnold diagram Fig. 7 whose frequency mismatch axis is expressed in $\delta/\Gamma_p$ units and the driving amplitude axis in $\Delta\Omega_S/\Gamma_p$ units. With this specific normalization, it is possible to compare different hysteretic ranges coming from forcings of different natures (use of $\Delta\Omega_S$ instead of $a_g$) as well as to verify the dependence of Eq. 10 on the operating power $p_0$ (axis rescaled with $1/\Gamma_p$).

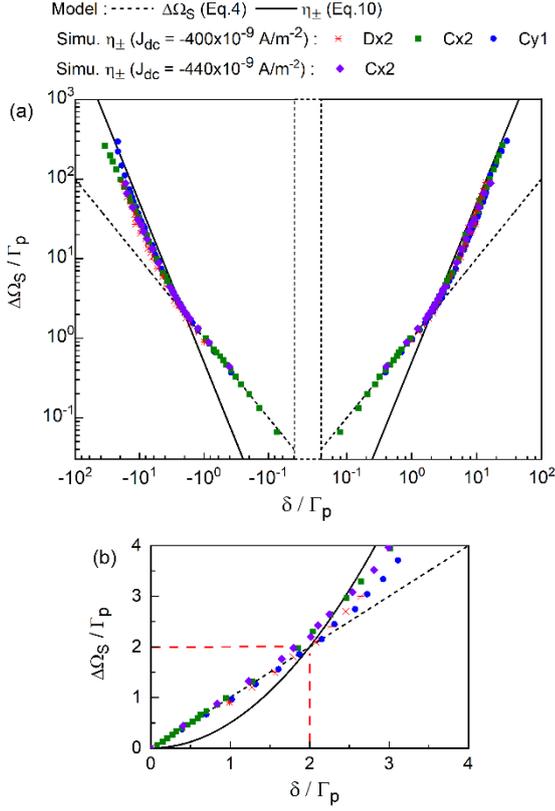

*Fig. 7. (a) Arnold tongue diagram with y-axis rescaled in $\Delta\Omega_S/\Gamma_p$ and x-axis rescaled in $\delta/\Gamma_p$. Lines, model predictions. Symbols, numerical results in case of the forcings "Dx2," "Cx2" and "Cy1" for different $J_{dc}$. At $J_{dc} = -400 \times 10^9 A.m^{-2}$: $p_0 = 0.093$ and $\Gamma_p(p_0) = 21.2\ MHz$. At $J_{dc} = -440 \times 10^9 A.m^{-2}$, $p_0 = 0.210$ and $\Gamma_p(p_0) = 55.5\ MHz$. Note in case of "Dx2" (red stars) or "Cy1" (blue circle), the locking range asymmetry can be neglected because of the weak driving amplitude. (b) Restricted view in the right plot of (a) with linear scaled axis around the critical threshold (red dashed line).*

One can identify two regimes depending on whether the driving amplitude is above or below a certain threshold. For driving amplitudes below the threshold, the hysteretic boundaries $\eta_\pm$ are identical to the locking range boundaries $\delta_\pm$ (dashed black line). In other words, there is no hysteresis in this regime. This regime have been first identified in Ref. [16] and holds as long as the locking range boundaries Eq. 6 remain inferior to the hysterical boundaries Eq. 10 ($\eta_+(a_g) < \delta_+(a_g)$). In particular, the condition $\eta_+(a_{g,th}) = \delta_+(a_{g,th})$, defining the driving amplitude threshold $a_{g,th}$, can be written in the general form : $\Delta\Omega_S = 2\Gamma_p$ if one neglects the locking range asymmetry ($\delta_\pm \approx \pm\Delta\Omega_S$) because of the weak driving amplitude, Fig 7b. At the opposite, when the driving amplitude exceeds the threshold, the hysteretic boundaries $\eta_\pm$ of all the investigated forcings begin to follow the relation Eq. 10. In this regime, the locking range becomes hysteretic. Interestingly, they continue to do so, over a wide range of driving amplitude corresponding to magnetic fields of the $10^3\ A.m^{-1}$ order and hysteretic ranges of the GHz order. We emphasize that Eq. 10 is verified by different cases (i) when the forcings are applied to the same free running conditions, i.e the same power $p_0$: for example, "Cy1" (blue circle) and "Dx2" (red stars) at $J_{dc} = -400 \times 10^9\ A.m^{-2}$, and (ii) identical forcings applied to different free running states $p_0$: for example, "Cx2" at $J_{dc} = -400 \times 10^9\ A.m^{-2}$ (green square) and at $J_{dc} = -440 \times 10^9\ A.m^{-2}$ (violet diamond), hence justifying the dependence on the oscillation power $p$ of Eq. 10.

However, we also note some discrepancies with the model. For example, the agreement between the simulation and the model is better for the positive hysteretic branch $\eta_+$ than for the negative one $\eta_-$, due to a slight asymmetry not predicted by the model. In Fig. 7a we see that the error for the negative branch increases progressively as the driving amplitude increases, while remaining quite low as it reaches only 15% in the worst case for the configuration "Cx2" at $J_{dc} = -400 \times 10^9\ A.m^{-2}$ and $H_g = 9549\ A.m^{-1}$. The origin of this asymmetry is difficult to state. However, since the asymmetry is quantitatively the same for all the forcings investigated (Fig. 7a), the origin should be at least independent on the forcings. A possible explanation is given in Ref. [13] and could be due to the high order dependence of the free running frequency with the oscillation power of the STNO ($\omega(p) = \omega(0) + \omega_{NL}(p)$ with $\omega_{NL}^{(k)}(p) \neq 0$ if $k \geq 2$).

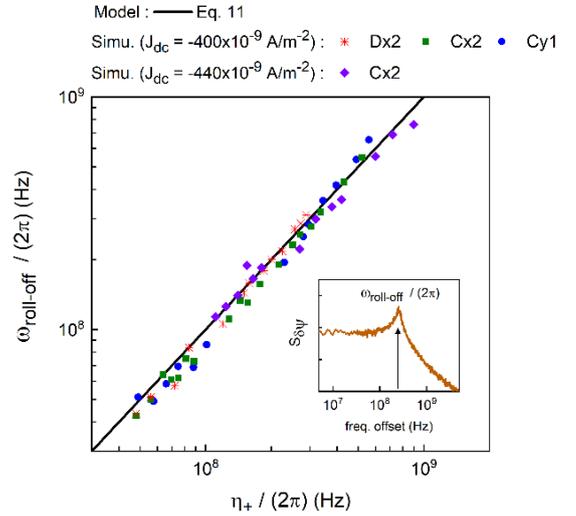

*Fig. 8. Resonant frequency versus positive hysteretic boundary. Line, model prediction. Symbols, numerical results in case of the forcings "Dx2," "Cx2" and "Cy1" of Fig. 7. Inset: Example of a numerical power spectral density of the phase fluctuations at $T = 1K$ for "Cx2" at $J_{dc} = -400 \times 10^9 A.m^2$, $H_g = 795\ A.m^{-1}$, see [32] for more details.*

This good agreement with the macrospin simulations in case of different forcings and range of driving amplitudes is a strong evidence in favor of Eq. 10. Moreover, it turns out that if Eq. 10 is correct, the hysteretic boundaries would be precisely the expression of the resonant angular frequency of the forced system, at zero detuning, calculated in the approximation $|\nu \varepsilon_1/\varepsilon_2| \gg 1$ Therefore, another way to express the hysteretic boundaries is (Appendix C):

$$\eta_\pm = \pm \omega_{roll-off} \quad (11)$$

With $\omega_{roll-off} = |\lambda_\pm|$, and $\lambda_\pm$ the eigenvalues of the Jacobian $\mathcal{J}(p, \psi)$ Eq. 8 taken at zero detuning ($p = p_0$ and $\psi = \pi/2$). More details on the properties of the oscillatory transient regime of the forced STNO are given in Refs. [18,32,33,34].

In Fig. 8, we have verified this relation by macrospin simulations for the three forcings "*Cx2*", "*Dx2*" and "*Cy1*" of Fig. 7. On the one hand, we have measured the resonant frequency $\omega_{roll-off}$ from the power spectral density of the phase fluctuations of the STNO at $T = 1K$ (inset Fig. 8). On the other hand, we have measured the positive hysteretic boundary $\eta_+$ for the same forcing configuration and driving amplitude. As a result, the comparison gives a 1 to 1 ratio between the two quantities, valid even for hysteretic boundaries of the GHz order and therefore justifying the equality.

Note that Eq. 11 is also verified for various types of forcings, whatever the nature of the signal, the harmonic orders, or the free running powers where they are applied, so that one might be tempted to say that Eq. 11 holds also for any other forcings and nonisochronous oscillators providing that the condition $|\nu \varepsilon_1/\varepsilon_2| \gg 1$ is respected. Such relation gives interesting perspectives for experiments or theory, because it allows to estimate the hysteretic boundaries $\eta_\pm$ only through the phase noise measurement at zero detuning, which avoid complicate scans of the generator frequency in different directions.

# 6. ARNOLD TONGUE DIAGRAMS AND HIGH DRIVING AMPLITUDE REGIMES

In this last part, we summarize the main features of the forcings "*Dx2*", "*Cx2*", "*Cy1*" seen in Part. 3-5 and confront the model with the simulation for a larger range of driving amplitudes. In particular, we shall see that the relative position in the Arnold tongue diagram between the locking range and hysteretic boundaries is critical to predict the dynamic of the STNO.

We turn to the Arnold tongue diagrams in terms of the power shift $\Delta p$ Fig. 9(a,c,e) and the frequency mismatch $\delta$ Fig. 9(b,d,f). The analysis of Fig. 9 is simplified if one notes that the power and frequency diagrams are related through $\Delta p \approx \delta/nN$ and symmetrically reversed because of the negative non linearity ($N < 0$) for the case of the IP-STNO. As a result, one should consider the power locking range boundaries $\Delta p_+$ ($\Delta p_-$) and the frequency locking range boundaries $\delta_-$ ($\Delta p_+$) to be equivalent. This way of viewing the Arnold tongue diagrams may be quite complicated at first sight, but it can be interesting specially to detect higher order bifurcation scenarios that break the relation $\Delta p \approx \delta/nN$ and the symmetry between

the power and the frequency of the oscillator. In the following each forcing is addressed successively:

*Case "Dx2"*

We start by focusing on the forcing "*Dx2*" Fig. 9(a,b) and first address the range of driving amplitudes $J_g/J_{dc} < 1.5$. Fig. 9(a,b) confirms all the predictions made by the model Part. 3-5. In particular, we see that the outer power locking range boundaries $\Delta p_\pm$ shift correctly towards positive power mismatch $\Delta p$ ($\varepsilon_1^{(1)}(p) > 0$) and the frequency boundaries $\delta_\pm$ to the negative frequency mismatch $\delta$ with the increase of the driving amplitude $J_g/J_{dc}$. Numerical simulations give $A_s = +6\%$ at $J_g/J_{dc} = 0.1$ and $A_s = +20\%$ at $J_g/J_{dc} = 0.3$. Moreover, we see that the expansion of the locking range boundaries $\Delta p_-$ and $\delta_+$ with the increase of $J_g/J_{dc}$ is not restricted due to the TB effect because on one hand the power threshold is $p_{TB} = 0$ ($\Delta p_{TB} = -p_0$) and on the other hand $\Delta p_- = -p_0$ only in the asymptotic limit $J_g/J_{dc} \gg 1$, see Eq. 6. Due to the asymmetry behavior, the opposite boundaries $\Delta p_+$ and $\delta_-$ drastically increases with small changes of the driving amplitude $J_g/J_{dc}$ Fig. 9(a,b). In particular, they are the strongest at $J_g/J_{dc} = 1.5$ with $\delta_-/2pi = -5.9 GHz$. Such large locking range is possible in STNOs due to a unique combination of high nonisochronicity and a strong dependence of the forcing on the oscillation power.

However above $J_g/J_{dc} = 1.5$, we see that the extension of $\Delta p_+$ and $\delta_-$ is limited or even decrease with the driving amplitude Fig. 9(a,b), which is in contradiction with the model. This scenario is referred as (S1). At such high level of oscillation power $p \approx 0.47$ (close to transition of the out plane precession mode), the spin wave formalism is less relevant so that it is difficult to explain this effect in the scope of the model. However, note that the decrease is qualitatively the same for $\Delta p_+$ and $\delta_-$ so that even in this regime of high driving amplitude, the relation $\delta \approx nN\Delta p$ continues to hold.

*Case "Cx2", "Cy1"*

For the magnetic field forcings "*Cx2*" and "*Cy1*", we see a good agreement between the model and the simulation specifically in the range $H_g < 10^3 A/m$. For "*Cx2*", the power and frequency boundaries $\Delta p_\pm$, $\delta_\pm$ remain symmetric with the driving field $H_g$ due to $\varepsilon_1^{(1)}(p) = 0$, Fig. 9(c,d). In contrast for "Cy1", the boundaries $\Delta p_\pm$ shift to the negative power mismatch $\Delta p$ Fig. 9(e,f) ($\varepsilon_1^{(1)}(p) < 0$) and the boundaries $\delta_\pm$ to the positive frequency mismatch $\delta$. A numerical simulation gives $A_s = -12\%$ at $H_g = 47.7 \, A/m$. Note that in case of "*Cx2*" and "*Cy1*", the boundary $\Delta p_-$ is mainly restricted due to the Taken-Bogdanov effect, so that it is difficult to observe the locking range asymmetry, but one can see it clearly for the opposite boundary $\Delta p_+$. Contrary to "*Dx2*", the forcings "*Cx2*" and "*Cy1*" undergo a TB bifurcation for a non-zero value of the threshold power $p_{TB} = p_0/2$ ($\Delta p_{TB} = -p_0/2$) which induces the restriction of the locking range. For "*Cx2*", the negative power bound $\Delta p_-$ becomes fixed at $H_g \approx 310\ A/m$ with the value $\Delta p_{TB} \approx -0.043 \approx -p_0/2$ as the driving field $H_g$ increases, Fig. 9(c). In a similar way, for "Cy1", $\Delta p_-$ becomes fixed at $H_g = 100\ A/m$ with the value $\Delta p_{TB} \approx -p_0/2$, Fig. 9(e). We see that $\Delta p_-$ remains fixed whatever the value of the driving field $H_g$, which confirms that $\Delta p_{TB}$ is independent of the driving field $H_g$. Regarding the behavior in frequency for "Cx2" "Cy1" Fig. 9(d,f), the increase of the positive frequency bound $\delta_+$ with $H_g$ is first restricted as predicted by the model due to the TB threshold but unexpectedly starts rising again when the TB boundary crosses the hysteretic boundary. For example, for "Cx2", the positive frequency bound $\delta_+$ becomes fixed at $H_g \approx 310\ A/m$ with the value $\delta_{TB} \approx 495\ MHz \approx -Np_0$ and start rising again for $H_g > 1500\ A/m$, see Fig. 9(f).

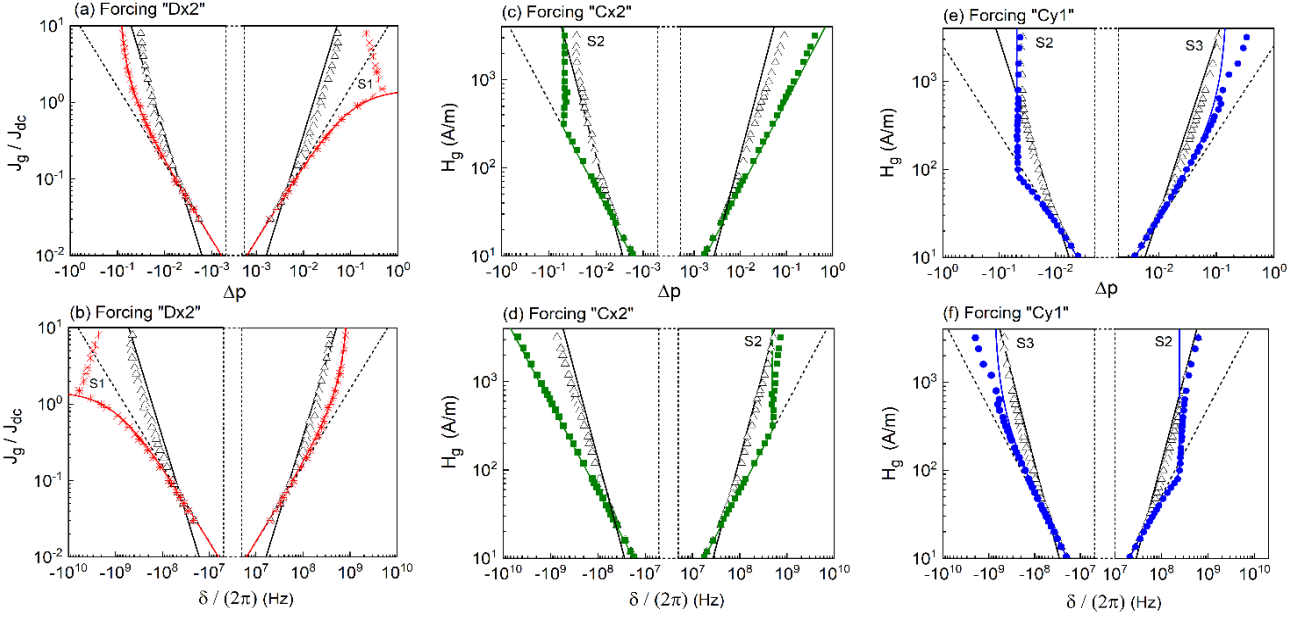

*Fig. 9. Arnold tongue diagrams of the in-plane uniform STNO and the forcings defined in Part 2, operated at $J_{dc} = -400 \times 10^9 A/m^2$ ($p_0 = 0.093$). Forcing cases: "Dx2", (a,b), "Cx2" (c,d) and 'Cy1'. (e,f). Power diagrams (a,c,e). Frequency diagrams (b,d,f). Numerical simulation: open triangle, hysteretic boundaries, symbols (star, square and circle), locking range boundaries. Model: dashed black line, Adler boundaries $\Delta\Omega_S$ Eq.4 or $\Delta\Omega_S/(nN)$. Colored line, locking range boundaries $\delta_\pm$ or $\delta_\pm/(nN)$ Eq.6 and TB bifurcation. Black line, hysteretic boundaries $\eta_\pm$ or $\eta_\pm/(nN)$ Eq.10.*

In particular, this behavior is an example of another higher order bifurcation scenario (S2) where model and simulation disagree. While the negative power bound $\Delta p_-$ remains fixed at the TB bifurcation threshold whatever the driving field, the corresponding positive frequency bound $\delta_+$ may increase. It is observed for the two magnetic field forcings "Cx2", "Cy1" and is triggered as soon as the positive hysteretic bound $\eta_+$ comes close to $\delta_+$ in the Arnold diagram in frequency Fig. 9(d,f). In this regime, one can explain the deviation of the numerical positive frequency bound $\delta_+$ by looking at the frequency vs. power dependence of the oscillator. If the saturation is observed for the power bound $\Delta p_-$ and not observed for the frequency bound $\delta_+$, one must conclude that the relation $\delta \approx nN$ is no longer valid. Actually, this is confirmed in Fig. 10 where we have plotted the frequency vs. power dependence for the forcing "Cy1" at low driving field $H_g = 218\ A/m$ and at high driving field $H_g = 3143\ A/m$. For the low value of the driving field, the frequency vs. power dependence clearly follows the same relation as in free running regime. The minimum power is $p_{TB} \approx p_0/2$ as predicted by the model. In contrast for high value of the driving signal, the frequency vs. power dependence deviates from the free running relation near the threshold power $p_{TB}$. In other words, the deviation between the numerical and predicted frequency bound $\delta_+$ is possible only at the expense of a modification of the power shape solution.

Besides, the forcing "Cy1" presents a last bifurcation scenario (S3) that challenge once more the model. In Fig. 9(e,f), we see that the increase of $\Delta p_+$ and $\delta_-$ with the driving field $H_g$ is limited also, except that this time, the cause is the asymmetry of the forcing. On the other hand, the hysterical bound $\eta_-$ growths with the driving field without any restriction so that the interaction with the locking range boundary $\delta_-$ is inevitable Fig. 9(f). For the locking range bound arising from a Taken-Bogdanov mechanism, we have previously observed that the hysterical bound affects much more the frequency locking range bound than the power bound. In contrast, we see for the locking range bound arising from a Saddle-node mechanism that a different mechanism takes place Fig. 9(e,f) where the hysterical bound affects equally the frequency bound $\delta_-$ and the power bound $\Delta p_+$. In terms of frequency vs. power dependence, we see in Fig. 10 for the case $H_g = 3143\ A/m$, that the hysterical $\eta_-$ bound repels simultaneously the frequency bound $\delta_-$ and the power bound $\Delta p_+$ according to the free running relation, which contrast with the opposite bounds $\delta_+$ and $\Delta p_-$ as discussed previously. Compared to the two others forcings, this forcing "Cy1" is interesting because it illustrates two different ways of interaction between the hysterical bounds and the locking range bounds, which depend on the bifurcation type involved.

We end this part by pointing out that all the discussed features regarding the hysteresis effect are fully reproducible by the general oscillator equation Eq. 2. In particular, this fact is not obvious for the prediction of the hysteretic boundaries since Eq. 9 is an intuitive estimation rather than a rigorous prove. Furthermore, the general oscillator equation can also describe finer effects of the dynamic observed in macrospin simulation and previously discussed, such as: (i) the slight asymmetry of the hysteretic branches, (ii) the high order bifurcation scenario (S2-3) that show the perturbation of the locking range boundaries by the hysteretic boundaries as the driving amplitude increases. Therefore, if there is a need to go beyond the calculations presented in this report, we find that the general oscillator equation Eq. 2 can be also a reliable and useful tool, easier to manipulate than the full LLGS equation, able to describe finer effects of the dynamic.

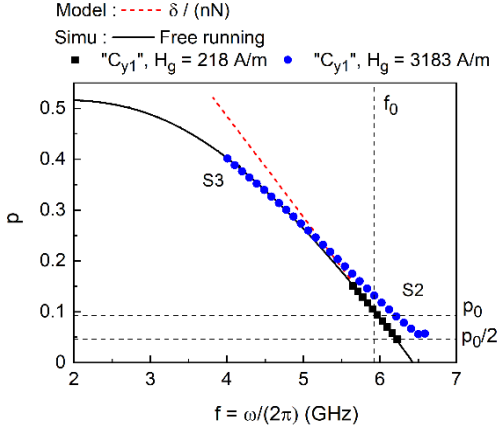

*Fig 10. Stationary power solution in case of the forcing "Cy1" operated at $J_{dc} = -400 \times 10^9 A/m^2$ or $p_0 = 0.093$ for two values of driving field $H_g = 218\ A/m$ (Blue circles) and $H_g = 3143\ A/m$ (black squares). Numerical power vs. frequency dependance in the free regime (black solid line), theoretical relation $\delta = nN\Delta p$ (red dashed line).*

## 6. CONCLUSION

The bifurcation diagram of injection locked STNOs presents several complications compared to that of isochronous auto-oscillator with the emergence in the locking range of different types of asymmetries as well as hysteresis effects. However, we have shown that when considering the limit of strong nonisochronicity (and/or forced by a power forcing), these effects can be predicted and discussed in a rather general way within the auto oscillator equation and given the properties of the STNO and the forcings. For example, we have shown that the forcing function is fully related to (i) the locking range asymmetry arising from Saddle-Node bifurcation, (ii) the locking range asymmetry arising from Taken-Bogdanov bifurcation and its power threshold and (iii) the frequency hysteretic range which is related to the transient regime of the system. Moreover, we have shown that the relative positions between all the bifurcation lines in the bifurcation diagram is critical to predict the dynamic of SNTOs and the validity of the model in high driving amplitudes regimes

It is possible that certain results, such as the prediction of the locking range boundaries, are difficult to used and verified, given the high thermal fluctuations encountered in experiments. However, it remains that the description of the hysteretic range boundaries as well as their interactions with other bifurcation lines of the system is of considerable interest, since for this range of mismatch frequencies, the locking effect is stable and maximized as wanted for applications. Such efforts strengthen our understanding of synchronized STNOs, especially strongly nonisochronous ones, and make them realistic candidates for the development of novel *rf* applications.

## ACKNOWLEDGMENT

J. H. acknowledges financial support from DGA. J. H. would like to thank Julie Chapelain for her support during the preparation of the manuscript.

## APPENDIX A : CONVERSION TO MX MY MZ COORDINATES

The Hamiltonian spin wave formalism consists in transforming the dynamic of the oscillator from the real space variable $\boldsymbol{m}(t) = (m_x, m_y, m_z)$ to complex valued variables $a(t)$, $b(t)$ and $c(t) = \sqrt{p}e^{i\phi(t)}$ with the oscillation power $p$ and the free running phase $\phi(t)$. The relation between $c(t)$ and the complex variable $d(t)$ used in this report is given by : $d(t) = c(t)\exp\left(-\frac{i\omega_g t}{n}\right)$.

The trajectory orbit $p = cste$ in $c$ variable can be express in the real space by:

$$m_x = 1 + 2p\left(\frac{\mathcal{B}}{\mathcal{A}}\cos(2\phi) - 1\right) \quad (A1)$$

$$m_y = \cos(\phi)\sqrt{2p\left(1 - p + \frac{\mathcal{B}}{\mathcal{A}}p\cos(2\phi)\right)}\left(\sqrt{1 + \mathcal{G}\left(\frac{\mathcal{B}}{\mathcal{A}}\right)} - Sign\left(\frac{\mathcal{B}}{\mathcal{A}}\right)\sqrt{1 + \mathcal{G}\left(\frac{\mathcal{B}}{\mathcal{A}}\right)}\right) \quad (A2)$$

$$m_z = -\sin(\phi)\sqrt{2p\left(1 - p + \frac{\mathcal{B}}{\mathcal{A}}p\cos(2\phi)\right)}\left(\sqrt{1 + \mathcal{G}\left(\frac{\mathcal{B}}{\mathcal{A}}\right)} + Sign\left(\frac{\mathcal{B}}{\mathcal{A}}\right)\sqrt{1 + \mathcal{G}\left(\frac{\mathcal{B}}{\mathcal{A}}\right)}\right) \quad (A3)$$

With the function $\mathcal{G}(x) = \sqrt{1-x^2}$, the ellipticity parameter $-1 \leq \mathcal{B}/\mathcal{A} \leq 1$ and $0 \leq \phi \leq 2\pi$. Some orbits are illustrated in Fig. A1.

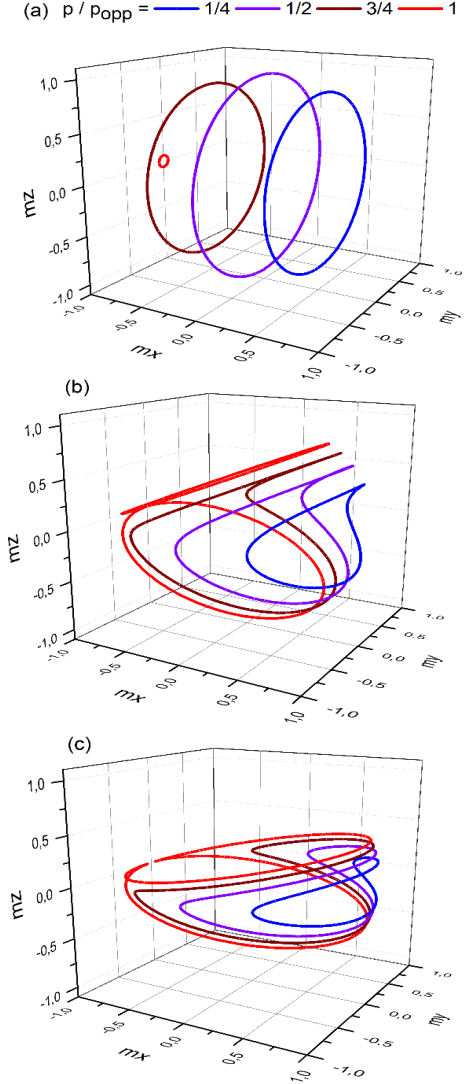

*Fig. A1. Theoretical orbits represented in the real space oriented along the $\mathbf{u}_x$ axis for different ellipticities $\mathcal{B}/\mathcal{A}$ and oscillation power $p$. (a) $\mathcal{B}/\mathcal{A} = 0$, (b) $\mathcal{B}/\mathcal{A} = -2/3$ and (c) $\mathcal{B}/\mathcal{A} \approx -0.91$ (studied IP-STNO). The power $p_{opp} = 1/(1+|\mathcal{B}/\mathcal{A}|)$ delimit the in-plane precession mode from the out of plane precession mode.*

## APPENDIX B : TAKEN-BOGDANOV BIFURCATION

The dynamical system Eq. 2 is resumed :

$$\dot{p} = -2Z(p) + 2p\varepsilon_1(p)a_g\cos(\psi) \quad (B1.1)$$

$$\dot{\psi} = n\omega(p) - \omega_g + n\varepsilon_2(p)a_g\sin(\psi) \quad (B1.2)$$

To simplify the calculations, we have set : $\theta = 0$ and $Z(p) = p\Gamma(p)$.

On the first hand, the stationary states of Eq. B1 must verify the equilibrium conditions $\dot{p} = 0$ and $\dot{\psi} = 0$, which rewrites :

$$\cos(\psi) = \frac{Z(p)}{pa_g\varepsilon_1(p)} \quad (B2.1)$$

$$\sin(\psi) = \frac{-n\omega(p) + \omega_g}{na_g\varepsilon_2(p)} \quad (B2.2)$$

On the other hand, the Jacobian matrix $\mathcal{J}$ of Eq. B1 evaluated at the stationary point $(p,\psi)$ is:

$$\mathcal{J}(p,\psi) = \begin{pmatrix} -2Z^{(1)} + 2a_g(\varepsilon_1 + p\varepsilon_1^{(1)})\cos(\psi) & -2pa_g\varepsilon_1\sin(\psi) \\ nN + na_g\varepsilon_2^{(1)}\sin(\psi) & na_g\varepsilon_2\cos(\psi) \end{pmatrix} \quad (B3)$$

Now, we focus on the quantity $Tr(\mathcal{J}(p,\psi))$:

$$Tr(\mathcal{J}) = -2Z^{(1)} + 2a_g(\varepsilon_1 + p\varepsilon_1^{(1)})\cos(\psi) + na_g\varepsilon_2\cos(\psi) \quad (B4)$$

Removing $\psi$ with Eq. B2.1 leads to :

$$Tr(\mathcal{J}) = \frac{Z}{p}\left(-2\frac{Z^{(1)}p}{Z} + \frac{2(\varepsilon_1 + p\varepsilon_1^{(1)})}{\varepsilon_1} + \frac{n\varepsilon_2}{\varepsilon_1}\right) \quad (B5)$$

Note the useful relation :

$$\frac{Z^{(1)}(p)p}{Z(p)} = 1 + \frac{p}{\Delta p} \quad (B6)$$

with $\Delta p = p - p_0$ and $p_0$ the free running power.

Inserted Eq. B6 in Eq. B5 gives:

$$Tr(\mathcal{J}) = \frac{2Z}{p\Delta p}\left(-p + \frac{p\Delta p\varepsilon_1^{(1)}}{\varepsilon_1} + \frac{\Delta p n\varepsilon_2}{2\varepsilon_1}\right) \quad (B7)$$

The Taken-Bogdanov power $p_{TB}$ is defined by $Tr(\mathcal{J}(p_{TB})) = 0$ so that we obtain Eq. 8:

$$0 = -p_{TB} + p_{TB}(p_{TB} - p_0)\frac{\varepsilon_1^{(1)}(p_{TB})}{\varepsilon_1(p_{TB})} + (p_{TB} - p_0)\frac{n\varepsilon_2(p_{TB})}{2\varepsilon_1(p_{TB})} \quad (B8)$$

## APPENDIX C : TRANSIENT REGIME

The Jacobian matrix $\mathcal{J}(p,\psi)$ Eq. 8 or Eq.B3 with $\varepsilon_2 = 0$ is evaluated at zero detuning ($\delta = 0$, $p = p_0$, $\psi = \pi/2$) :

$$\mathcal{J}(p,\psi) = \begin{pmatrix} -2\Gamma_p & -2pa_g\varepsilon_1 \\ nN & 0 \end{pmatrix} \quad (C1)$$

The eigenvalues of the matrix are given by:

$$\lambda_\pm = -\Gamma_p\left(1 \pm \sqrt{1 - \frac{2\Delta\Omega_S}{\Gamma_p}}\right) \quad (C2)$$

The transient regime becomes oscillatory for the condition $\Delta\Omega_S > \Gamma_p/2$ with the complex eigenvalues:

$$\lambda_\pm = -\Gamma_p \left(1 \pm i \sqrt{\frac{2\Delta\Omega_S}{\Gamma_p} - 1}\right) \quad (C3)$$

Then, one can define the corresponding resonant frequency of the transient response:

$$\omega_{roll-off} = |\lambda_\pm| = \sqrt{2\Gamma_p \Delta\Omega_S} \quad (C4)$$